\documentclass[sigconf ]{acmart}
\usepackage{multirow}
\usepackage{algorithm}
\usepackage{algorithmic}
\usepackage{hyperref}
\usepackage{xcolor}
\usepackage{enumitem}
\usepackage{amsmath,amsfonts,bm}
\usepackage{titlesec}
\usepackage{amsmath}
\usepackage{bm}
\usepackage{enumitem}
\usepackage{booktabs}

\usepackage{subfig}

\AtBeginDocument{%
  }

\setcopyright{acmlicensed}
\copyrightyear{2018}
\acmYear{2018}
\acmDOI{XXXXXXX.XXXXXXX}

\acmConference[Conference acronym 'XX]{Make sure to enter the correct
  conference title from your rights confirmation emai}{June 03--05,
  2018}{Woodstock, NY}
\acmISBN{978-1-4503-XXXX-X/18/06}

\begin{document}

\title{HierLLM: Hierarchical Large Language Model for Question Recommendation}






\author{Yuxuan Liu}
\affiliation{%
  \institution{Jilin University}
  \country{Jilin Changchun}}
\email{liuyuxuan23@mails.jlu.edu.cn}

\author{Haipeng Liu}
\affiliation{%
  \institution{Jilin University}
  \country{Jilin Changchun}}
\email{liuhp22@mails.jlu.edu.cn}

\author{Ting Long}
\affiliation{%
  \institution{Jilin University}
  \country{Jilin Changchun}}
\email{longting@jlu.edu.cn}



\renewcommand{\shortauthors}{Trovato et al.}


\begin{abstract}
  Question recommendation is a task that sequentially recommends questions for students to enhance their learning efficiency. That is, given the learning history and learning target of a student, a question recommender is supposed to select the question that will bring the most improvement for students. Previous methods typically model the question recommendation as a sequential decision-making problem, estimating students' learning state with the learning history, and feeding the learning state with the learning target to a neural network to select the recommended question from a question set.
  However, previous methods are faced with two challenges: (1) learning history is unavailable in the cold start scenario, which makes the recommender generate inappropriate recommendations; (2) the size of the question set is much large, which makes it difficult for the recommender to select the best question precisely. To address the challenges, we propose a method called hierarchical large language model for question recommendation (HierLLM), which is a LLM-based hierarchical structure. The LLM-based structure enables HierLLM to tackle the cold start issue with the strong reasoning abilities of LLM. The hierarchical structure takes advantage of the fact that the number of concepts is significantly smaller than the number of questions, narrowing the range of selectable questions by first identifying the relevant concept for the to-recommend question, and then selecting the recommended question based on that concept. This hierarchical structure reduces the difficulty of the recommendation.To investigate the performance of HierLLM, we conduct extensive experiments, and the results demonstrate the outstanding performance of HierLLM.

\end{abstract}



\received{20 February 2007}
\received[revised]{12 March 2009}
\received[accepted]{5 June 2009}

\maketitle
\section{Introduction}\label{sec:intro}
Question recommendation is a task that recommends personalized questions for students to exercise, enabling them to efficiently make improvement in learning from a long-term perspective. Formally, given the learning history and learning target of an arbitrary student, a question recommender aims to sequentially recommend questions for the student.
As it is illustrated in Figure \ref{sec:intro}, the learning target of the student is mastering question with ID \( v_1 \), \( v_3 \) and \( v_9 \). Given the learning history of students' answering questions $\{q_1, q_2, ..., q_8\}$, the recommender is supposed to sequentially recommend questions for the student to make them master the learning targets efficiently.
Compared to traditional teaching methods, which typically employ standardized teaching plans and overlook individual differences among students, question recommender cater to the unique knowledge levels and learning targets of individual students, making students to grasp knowledge and achieve their learning targets more efficiently.

\begin{figure}[t]
\vspace{-10pt}
\centering
\includegraphics[width=1\columnwidth]{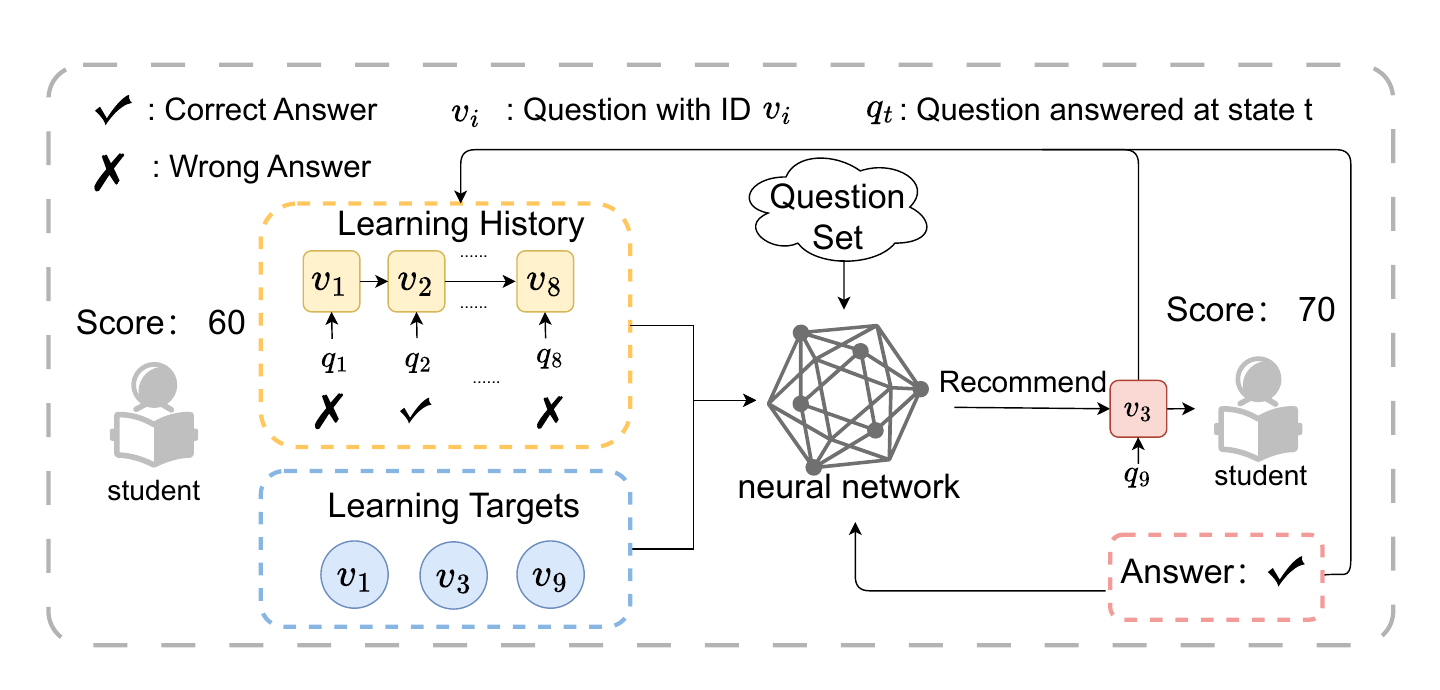} 
\vspace{-20pt}
\caption{Illustration of a recommender recommends questions based on learning history and learning targets. }
\vspace{-20pt}

\label{fig: intro}

\end{figure}

To learn an effective question recommender, previous methods commonly model question recommendation as a sequential decision-making problem and optimize the recommender with reinforcement learning (RL) algorithms \cite{chen2023set,li2023graph,liu2019exploiting}. Specifically, they first encode the learning history of the student as a vector to represent the student's learning state, and then they feed the learning state with learning target representation to a neural network to select the recommended question from a question set.
As shown in Figure \ref{fig: intro}, these methods estimate the learning state of the student according to their learning history spanning from question \( q_1 \) to \( q_8 \) first. Then, they feed the learning state and learning target $\{v_1, v_3, v_9\}$ to a neural network, select a question with ID \( v_3 \) from the question set, and recommend  \( v_3 \) to the student. By repeatedly estimating the latest learning state and recommending new questions until the end of the recommendation session, they obtain the feedback of the recommendation and optimize the recommender according to the feedback.

However, previous methods are faced with two challenges from the perspective of temporal and spatial: (1) temporal issue: previous methods are dependent on learning history to estimate the learning state of students. However, in cold start scenarios, the learning history is unavailable, resulting the inappropriate recommendation; (2) spatial issue: previous methods are required to select one question from the question set and recommend it to students. However, in real-world scenarios, the size of the question set is relatively large, reaching up to several thousand or even hundreds of thousands, which makes it difficult to select the beneficial question precisely from such a large decision space.

To address the limitations of previous methods, we propose a method called hierarchical large language model for question recommendation (HierLLM), which is a hierarchical structure built based on the large language model (LLM). 
Taking advantage of LLM in reasoning capabilities and extensive training knowledge\cite{pan2024unifying}, HierLLM can be feasible in tackling the cold start issue. 
Taking advantage of the fact that the number of concepts is significantly smaller than the number of questions, HierLLM leverages its hierarchical structure to narrow the range of selectable questions. By first identifying the relevant concept for the to-recommend question, and then selecting the recommended question based on that concept, HierLLM can significantly reduce the difficulty of the recommendation.
To evaluate the performance of HierLLM, we conducted extensive experiments, and the experiment results demonstrate that HierLLM reaches state-of-art performance. 

In summary, our paper makes the following contributions:
\vspace{-2pt}
\begin{itemize}[leftmargin=15pt]
     \item We propose a method called hierarchical large language model for question recommendation (HierLLM) to recommend personalized questions for students.
    
    \item HierLLM introduces a hierarchical LLMs for to address the issue of cold-start and large-size question sets in educational scenarios. 


    \item We evaluate our HierLLM through extensive experiments, and the experiment results demonstrate the effectiveness of our method.

\end{itemize}

\vspace{-10pt}
\section{Related Works}
\subsection{Personalized Question Recommendation}

Many outstanding approaches have been proposed to address the question recommendation task. Some works attempt to use general recommendation algorithms\cite{zhou2018personalized, liu2020research}, to recommend similar paths for students with similar learning target. Other works propose more targeted personalized question recommendation methods. For example, CSEAL\cite{liu2019exploiting} ensures logical question recommendations through a navigation algorithm on the knowledge structure and uses an actor-critic algorithm to dynamically update strategy parameters. SRC\cite{chen2023set} employs an attention mechanism to explore the correlation between questions and optimizes the model based on student feedback. GEHRL\cite{li2023graph} utilizes a hierarchical reinforcement learning method to plan students’ learning paths. Although previous methods have achieved certain results, they have overlooked the problem of cold start and overly large question sets. In our method, we incorporate a LLM-based hierarchical structure to address the limitations.

\vspace{-10pt}
\subsection{Application of LLM in Recommendations}
Recently, large language models (LLMs) have demonstrated impressive reasoning and decision-making capabilities across various tasks\cite{eigner2024determinants, zhang2024llm,ma2024towards}, and many works have explored the application of LLMs in recommendation systems. These works can be roughly grouped into two categories. The first category utilizes LLMs for feature enhancement. For instance, KAR\cite{xi2023towards}uses LLMs to acquire knowledge about user preferences and items and then inputs these representations into traditional recommendation models to enhance performance. ONCE\cite{liu2023first} enhances content recommendation systems by combining open-source and closed-source large language models. The second category directly utilizes LLMs for recommendations. For example, TALLRec\cite{bao2023tallrec} adopts the low-rank adaptation(LoRA)\cite{hu2021lora} method to fine-tune LLMs for recommendations. A-LLMRec\cite{kim2024large} aligns collaborative and textual knowledge with a frozen collaborative filtering system and uses the frozen LLM to integrate this knowledge during the recommendation phase. Given the powerful capabilities of LLMs, we explore personalized question recommendations based on LLMs.

\begin{figure*}[t]
\centering
\vspace{-10pt}  
\includegraphics[width=1.8\columnwidth]{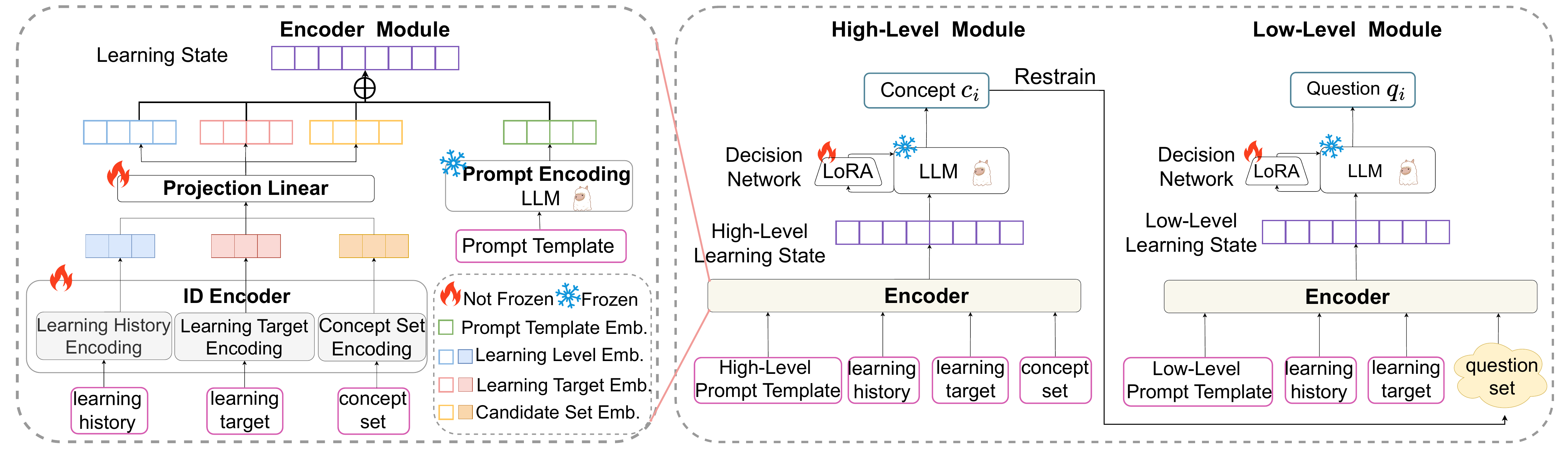}
\vspace{-10pt}
\caption
{The framework of HierLLM. HierLLM is a LLM-based hierarchical structure, which is composed of a high-level module and a low-level module (right side). The high-level module is responsible for recommending concept to narrow the recommendation space to a question candidate set, dereasing the difficulty of recommendation.The low-level module is responsible for selecting a question from the question candidate set.}
\vspace{-10pt}
\label{fig: pipeline}
\end{figure*}

\vspace{-8pt}
\section{Preliminaries}
\subsection{Terminology}
Given the set of concepts $\mathcal{C} = \{ c_1, c_2, \ldots, c_m \} $, it consists of $ m $ concepts (\textit{e.g.}, for a mathematical learning site, it might include addition, subtraction as a concept). 
Given the set of questions $ \mathcal{Q}$, it consists of $ n $ questions, whose ID are denoted as $ \{ v_1, v_2, \ldots, v_n \} $ (\textit{e.g.}, for a mathematical learning site, $123 + 563 = ?$ might be one of the questions).
Obviously, each concept can correspond to multiple questions, and each question can cover multiple concepts. For most online learning sites, it holds that $m \ll n$. For a student $ u $, their learning history at step t is denoted as $ \mathcal{H}_t = \{(q_1, y_1), (q_2, y_2), ..., (q_{t-1}, y_{t-1}\}$, where $ q_t$ is a question from $\mathcal{Q}$ represents the question learned at time $ t$, and
the ID of question $q_t$ is denoted as $v_t$. $ y_t $ indicates whether the answer to $q_t$ was correct. Specifically, $ y_t = 0 $ denotes the student incorrectly answer question $q_t$, while $ y_t = 1 $ denotes the student correctly answered question $q_t$.
The learning targets of the student $ u $ are represented by a group of question IDs $ \mathcal{G} = \{ v_i\}^{|\mathcal{G}|} $, $v_i$ indicate one of question the student aims to master after learning the recommended questions. 
To measure the improvement after the student learning the recommended questions, the learning effect is defined, following previous works\cite{chen2023set,liu2019exploiting,li2023graph}:
\vspace{-6pt}
\begin{equation} \label{eq:learn-effect}
    \Delta_u = \frac{E_{a} - E_{b}}{E_{max} - E_{b}},
\vspace{-5pt}    
\end{equation}
where $ E_{a} $ and $ E_{b} $ are the number of questions in learning target correctly answered by the student after and before learning the recommended questions, respectively, and $ E_{max} $ is the number of questions in learning target.
\vspace{-4pt}
\subsection{Problem formulation.}
In this paper, we aim to develop a question recommendation model that adaptively recommends questions to students, maximizing their learning effect ($ \Delta_u $) from a long-term perspective. 
We model the recommendation problem as a Hierarchical Markov Decision Process (HMDP), which is composed of high-level MDP and low-level MDP.
The high-level MDP predicts the relevant concept for the to-recommend question to narrow the range of selectable questions, while the low-level MDP selects the recommended question from those related to the predicted concept, reducing the difficulty of the recommendation. Specifically, the HMDP includes:

\begin{itemize}[left=0cm]
\vspace{-4pt}
\item \textbf{$\mathcal{S}_h$ / $\mathcal{S}_l$} denotes the set of states in high-level and low-level MPD correspondingly. Given step $t$, for $\textbf{s}_t^h \in \mathcal{S}_h$ represents a student's learning state in high-level MDP (denoted as high-level learning state). For $\textbf{s}_t^l \in \mathcal{S}_l$ represents a student's learning state in low-level MDP (denoted as low-level learning state). The obtaining of both high-level and low-level learning state will be discussed in Section \ref{sec:encoder}.

\item \textbf{$\mathcal{A}_h$ / $\mathcal{A}_l$} denotes the finite set of actions in high-level MDP and low-level MDP, and they denote the concept set $\mathcal{C}$ and the candidate questions related to the concept outputted from high-level MDP. Given a step $t$, the high-level decision policy $\pi_h$ selects a concept $c_t$ from concept set $\mathcal{C}$ based on high-level learning state, \textit{i.e.}, $\pi_h(c_t|\textbf{s}_t^h, \mathcal{C})$. The low-level decision policy $\pi_l$ selects a question $q_t$ from the question candidate set $\mathcal{Q}_t$, in which all the questions are related to concept $c_t$, \textit{i.e.}, $\pi_l(q_t|(\textbf{s}_t^l, \mathcal{Q}_t))$. $\pi_h$ and $\pi_l$ will be discussed in Section \ref{sec:method}.



\item \textbf{$\mathcal{P}_h$ / $\mathcal{P}_l$} denotes the transition probability of transiting learning states in high-level and low-level MDP. 

\item $\mathcal{R}$ denotes the instant reward (\textit{i.e.}, the instant improvement bring for students) of selecting concept $c_t$ on state $\textbf{s}_t^h$ and selecting concept $q_t$ on state $\textbf{s}_t^l$, \textit{i.e.}, $r(q_t | \textbf{s}_t^l, \mathcal{Q}_t(c_t|\textbf{s}_t^h, \mathcal{C}))$

\item Discount factor $\gamma$ is the discount factor and $\gamma \in [0, 1]$.
\end{itemize}

\vspace{-4pt}
Therefore, the learning of question recommendation is maximizing students' long-term learning effects, \textit{i.e.},
\vspace{-5pt}
\begin{equation} 
\mathop{\max}_{\pi_h, \pi_l} J(\pi_h, \pi_l) 
=  \mathop{\max}_{\pi_h, \pi_l}  \Delta_u = \mathop{\max}_{\pi_h, \pi_l} \frac{1}{T} \sum_{t=0}^{T} 
r(q_t | \textbf{s}_t^l, \mathcal{Q}_t(c_t|\textbf{s}_t^h, \mathcal{C})),
\vspace{-4pt}
\end{equation}
where $T$ denotes the total length of recommendation sessions.

\vspace{-8pt}
\section{Methodology} \label{sec:method}

\subsection{Overview}
To recommend personalized questions and address both temporal and spatial challenges, we propose a method called the Hierarchical Large Language Model for Question Recommendation (HierLLM). HierLLM is a hierarchical structure built based on the large language model (LLM) and composed of a high-level module and a low-level module. Both high-level and low-level include an encoder and a decision network as shown in Figure \ref{fig: pipeline}. 
At each time step, the raw data of a student's learning history $\mathcal{H}_t$, learning target $\mathcal{G}$,
the concept set $\mathcal{C}$, and the prompt template to integrate $\mathcal{H}_t$, $\mathcal{G}$ and $\mathcal{C}$ are fed to encoder.
Then, the encoder tansforms them into a vector to represent the student's high-level learning state ($\textbf{s}_t^h$) and feeds the learning state to the decision network ($\pi_h$) of the high-level module. Then, the decision network of the high-level module leverages the learning state to predict the relevant concept for the to-recommend question,
and feed the predicted concept to the low-level module. 
Next, the low-level module leverages the concept to filter the questions unrelated to it and obtain the question candidate set $\mathcal{Q}_t$. Subsequently, the low-level module feeds the  learning history $\mathcal{H}_t$, learning target $\mathcal{G}$, question candidate set $\mathcal{Q}_t$ and the prompt template to integrate $\mathcal{H}_t$, $\mathcal{G}$ and $\mathcal{Q}_t$ to the encoder.
The encoder encodes them
to represent the low-level learning state ($\textbf{s}_t^l$) and feeds the low-level learning state to the decision network ($\pi_l$) of the low-level module. Finally, the decision network selects a question from the question candidate set and recommends it to the student.
In the following, we will present the high-level module and low-level module in HierLLM framework first and then we will discuss the details of the encoder in both modules.

\subsection{The Framework} \label{sec:framework}
As we discussed previously, the framework of HierLLM is a hierarchical structure based on LLM, 
which adopts a high-level module and a low-level module to obtain the recommended question in a step-by-step manner. As the input of the low-level module is the output of the high-level module, we will discuss the high-level module first, and subsequently the low-level module.

\subsubsection{The High-level Module.}
Given the learning history $\mathcal{H}_t$, learning target $\mathcal{G}$, and the concept set $\mathcal{C}$,
the high-level module is responsible for identifying the relevant concept ($c_t$) for the recommended question, narrowing the range of selectable questions.
To achieve that, the high-level module first feeds the $\mathcal{H}_t$, $\mathcal{G}$, $\mathcal{C}$, and the prompt template to integrate them together into an encoder
to obtain the representation of the student's learning state. We conduct this implementation based on the conclusion of previous works \cite{zhang2023collm, yu2024ra}, integrating domain-specific information in recommendations (\textit{e.g.}, the learning history, learning target) with the general knowledge held by LLMs into the same space enhances recommendation performance. Then, the high-level module feeds the student's learning state into the decision network of the high-level module to predict the relevant concept for the to-recommend question. Specifically, the high-level module takes the following steps:

First, the high-level module creates a template as illustrated in Table \ref{prompt template}, which we denote as the high-level prompt ($\mathcal{P}_t^h$), to integrate the concept set, the student's learning history and learning target.

Then, the high-level module feeds the learning history $\mathcal{H}_t$, learning target $\mathcal{G}$, concept set $\mathcal{C}$ and the high-level prompt to the encoder to obtain the representation of student's high-level learning state.
\vspace{-2pt}
\begin{equation}
    \textbf{s}_t^h = \text{Enc}(\mathcal{H}_t, \mathcal{T}, \mathcal{C}, \mathcal{P}_t^h),
\end{equation}
where \( Enc(\cdot)\) denotes the encoder, which will discussed in \ref{sec:encoder}. $\textbf{s}_t^h$ represents the student's high-level learning state, and $\textbf{s}_t^h \in \mathbb{R}^{d_m}$, which $d_m$ is an integer.

Next, the learning state is then fed to the decision network to output the probability of selecting each concept:
\vspace{-4pt}
\begin{equation}
    \textbf{p}_t^c = \pi_h(\textbf{s}_t^h),
\end{equation}
where $\pi_h(\cdot)$ is implemented with a pretrained LLM.
We apply Llama2-7B \cite{touvron2023llama} in our implementation. $\textbf{p}_t^c$ denotes the probability of selecting each concept and $\textbf{p}_t^c \in \mathbb{R}^{|\mathcal{C}|}$. We obtain the relevant concept of the recommended question by sampling a concept according to $\textbf{p}_t^c$. That is:

\vspace{-12pt}
\begin{equation}
    c_t \sim \textbf{p}_t^c,
\end{equation}
here $c_t$ denotes the selected concept of the to-recommend question.

\subsubsection{The Low-level Module.}

Given the learning history $\mathcal{H}_t$, learning target $\mathcal{G}$, and the concept $c_t$ covered by the to-recommend question,
the low-level module aims to identify the questions that bring the most significant long-term improvement from the questions related to $c_t$.
The low-level module first filters the questions that are unrelated to $c_t$ and obtains the question candidate set $\mathcal{Q}_t$. Then, like the high-level module, the low-level module designs a template prompt to integrate $\mathcal{H}_t$, $\mathcal{G}$, and $\mathcal{Q}_t$ and feed them with the template prompt into an encoder
to obtain the student's low-level learning state. Then, the low-level module leverages the low-level learning state to predict the question that brings the most improvement for students. 

Specifically, the low-level module first filters the questions that are unrelated to the concept output by the high-level module ($c_t$) from the question set, and obtains the candidate question set $\mathcal{Q}_t$.
Then, it designs a template of the low-level prompt ($\mathcal{P}_t^l$) as illustrated in Table \ref{prompt template}, which integrates the question candidate set $\mathcal{Q}_t$, the student's learning history $\mathcal{H}_t$ and learning target $\mathcal{G}$, and feeds them into an encoder.

\vspace{-10pt}
\begin{equation}
    \textbf{s}_t^l = \text{Enc}(\mathcal{H}_t, \mathcal{G}, \mathcal{Q}_t, \mathcal{P}_t^l),
\end{equation}
where \(Enc(\cdot)\)  denotes the encoder, which will discussed in \ref{sec:encoder}. $\textbf{s}_t^l$ represents the student's low-level learning state, and $\textbf{s}_t^l \in \mathbb{R}^{d_m}$.

Next, the low-level encoder feeds the low-level learning state to the decision network of the low-level module, which is also a LLM, to predict the recommendation probability of the questions in the candidate set:
\vspace{-4pt}
\begin{equation}
    \textbf{p}_t^q = \pi_l(\textbf{s}_t^l),
\end{equation}
where where $\pi_l(\cdot)$ is also implemented Llama2-7B in our implementation. $\textbf{p}_t^q$ denotes the recommendation probability of the candidate questions and $\textbf{p}_t^q \in \mathbb{R}^{|\mathcal{Q}_t|}$. We obtain the question recommending to the student by sampling the question according to $\textbf{p}_t^q$. That is:
\vspace{-4pt}
\begin{equation}
    q_t \sim \textbf{p}_t^q,
\end{equation}
where $q_t$ denotes the question recommended to the student.

It is important to clarify that although both high-level and low-level modules have an encoder and a LLM-based decision network, they share the same structure but different parameters.

\begin{table}[t!]
\centering
\vspace{-10pt}
\caption{Prompt Template}\label{prompt template}%
\vspace{-10pt}
\begin{tabular}{@{}p{220pt}@{}} 
\toprule
\multicolumn{1}{c}{\textbf{Prompt template for the High-Level}} \\
\midrule
\raggedright \hspace{2em}\#\#\# Instruction: Based on students' learning history: [ID\textsubscript{learning history}], learning targets: [ID\textsubscript{targets}] and concepts: [ID\textsubscript{concepts}], Predict the next concept. \#\#\# Response: 
\hrule
\vspace{3pt}
\centering \textbf{Prompt template for the Low-Level} \\
\vspace{3pt}
\hrule
\raggedright \hspace{2em}
\#\#\# Instruction: Based on students' learning history: [ID\textsubscript{learning history}], learning targets: [ID\textsubscript{targets}], and questions: [ID\textsubscript{questions}], Predict the next question. \#\#\# Response: 
\hrule
\end{tabular}
\vspace{-6pt}
\end{table}

\subsection{The Encoder}\label{sec:encoder}

As we discussed in previously, the encoder is used to convert student learning history, learning target, question set (concept set) and the prompt into vectors, and integrate their vector representation into a same space to represent the student's learning state. In the following, we will discuss how the encoder encode them separately and how it integrates them into the same vector space.

\textbf{Question candidate / Concept set encoding.}
Given the concept (or the question candidate) set, we aim to encode the set into a vector representation. Since both the concept and question set are the set, for the convenience of discussion, let us denote a set as $\mathcal{X} = \{x_1, x_2, ..., x_k\}$. $\mathcal{X}$ denotes the concept set or question candidate set. If $\mathcal{X}$ denotes the concept set, $x_i \in \mathcal{X}$ denotes one concept. If $\mathcal{X}$ denotes the set of question candidates, $x_i \in \mathcal{X}$ denotes one question candidate.

For $x_i \in \mathcal{X}$, we first encode it with one-hot encoding, and apply the attention mechanism to obtain its attentive representation.

\begin{equation}
\begin{aligned}
    \mathbf{e_{i}^o} &= \text{one-hot}(x_i), \\
    \mathbf{e_{i}^a} &= \text{Attn}(\mathbf{e}_i^o, \mathbf{E}_i^o),
\end{aligned}
\end{equation}
where $\mathbf{e_{i}^o}  \in  \mathbb{R}^{|\mathcal{X}|} $ is the one-hot encoding of $x_i$, and $\mathbf{E}_i^o \in  \mathbb{R}^{|\mathcal{X}|\times |\mathcal{X}|}$ is obtained by stacking the one-hot encoding of the elements in $\mathcal{X}$ into a matrix. one-hot($\cdot$) denotes the one-hot encoding, and Attn($\cdot$) denotes the attention mechanism.

Next, we obtain the representation of set $\mathcal{X}$ by:
\begin{equation}
    \mathbf{e}_t = Mean([f_o(\mathbf{E}^o) ; \mathbf{E}^a]), 
\end{equation}
where $Mean(\cdot)$ denotes the average pooling.  $([\cdot;\cdot)$ represents the concatenation operation.  \( f_o(\cdot)\) is the linear function. $\mathbf{E}^a \in \mathbb{R}^{|\mathcal{X}| \times {d_a}} $ is obtained by stacking the attentive representation of the elements in $\mathcal{X}$. $d_a$ is the dimension of the attentive representation. If $\mathcal{X}$ represents the concept set, we denote $\mathbf{e}_t \in \mathbb{R}^{2d_a}$ as $\mathbf{e}_t^c$ to denote the vector representation of concept set. If $\mathcal{X}$ represents the question candidate set, we denote $\mathbf{e}_t$ as $\mathbf{e}_t^q$ to represent the vector representation of question candidate set.  

\textbf{Learning target encoding.}
Like the question candidate or concept set encoding, we represent the learning target by averaging the attentive encoding in the set of the learning target. That is:
\vspace{-4pt}
\begin{equation}
    \mathbf{g}_t = Mean(\mathbf{E}_g^a), 
\end{equation}
where $\mathbf{E}_g^a \in \mathbb{R}^{|\mathcal{G}| \times 2d_a}$ is obtained by stacking the attentive representation of the questions in the learning target.

\textbf{Learning history encoding.}
Given the student's learning history 
$\mathcal{H}_t = \{(q_1, y_1), (q_2, y_2), ..., (q_t, y_t)\}$, we encode each record $(q_i, y_i)$ by 
\vspace{-4pt}
\begin{equation}
\textbf{z}_i = f_z \left([\textbf{e}_i; \ y_i]\right),
\end{equation}
where $f_z(\cdot)$ denotes a linear function.
$\textbf{e}_i \in \mathbb{R}^{|\mathcal{X}|}$ denotes the one-hot encoding of question $q_i$. $y_i$ denotes the correctness of the answer to question $q_i$. $\textbf{z}_i \in \mathbb{R}^{d_z} $, and  $d_z$ is the dimension of record representation.

Then, we initialize the representation of the learning history as $\textbf{h}_0$, which is a zero vector, \textit{i.e.}, $\textbf{h}_0 = \bm{\vec{0}}$, and feed $\textbf{z}_i$ to an Recurrent Neural Network to obtain the representation of the learning history:
\begin{equation} \label{eq:learn-his}
\textbf{h}_i = \text{RNN}(\textbf{z}_i, \textbf{h}_{i-1}), \text{for $1 \leq i \leq t$},
\vspace{-4pt}
\end{equation}
where RNN($\cdot$) denotes the RNN, and we implement it with Long Short-Term Memory (LSTM)\cite{graves2012long}. By applying Eq.(\ref{eq:learn-his}) $t$ times, we obtain $\textbf{h}_t \in \mathbb{R}^{d_h}$ and use the $\textbf{h}_t$ to represent the student's learning history. $d_h$ is the dimension of the learning history.

\textbf{Prompt encoding.}
As we discussed previously, the prompts are the text. Therefore, we encode both high-level prompt and low-level prompt with a pretrained LLM, That is:
\vspace{-4pt}
\begin{equation}
    \textbf{m}_t = PLLM(\mathcal{P}_t),
\end{equation}
where $\mathcal{P}_t$ denotes the high-level prompt $\mathcal{P}_t^h$ or low-level prompt $\mathcal{P}_t^l$ as we discussed in section \ref{sec:framework}. $\textbf{m}_t \in \mathbb{R}^{d_m}$ denotes the vector representation of the prompt. If it represents the high-level prompt, we denote it as $\textbf{m}_t^h$. If it represents the low-level prompt, we denote it as $\textbf{m}_t^l$

Finally, we sum the vector representation of learning history, learning target, concept set and high-level prompt to obtain the high-level learning state of the student:
\begin{equation}
    \textbf{s}_t^h = f_{p_1}(\textbf{h}_t) + f_{p_2}(\textbf{g}_t)+ f_{p_3}(\textbf{e}_t^c) + \textbf{m}_t^h,
\end{equation}

we sum the vector representation of learning history, learning target, question candidate set and low-level prompt to obtain the low-level learning state of the student:
\begin{equation}
    \textbf{s}_t^l = f_{p_1}(\textbf{h}_t) + f_{p_2}(\textbf{g}_t) + f_{p_4}(\textbf{e}_t^q) + \textbf{m}_t^l , 
\end{equation}
where \( f_{p_1}(\cdot) \), \( f_{p_2}(\cdot) \), \( f_{p_3}(\cdot) \), and \( f_{p_4}(\cdot) \) denote projection layers, which are implemented as linear functions.





\vspace{-8pt}
\subsection{Optimization}
To enable personalized question recommendation capabilities with the LLM, we froze the parameters of the pre-trained model and applied low-rank adaptation(LoRA)\cite{hu2021lora} for fine-tuning. Specifically, we defined three learning targets to enhance the learning of the high-level module, the low-level module, and the history representation in both modules.

Since the high-level module is defined as MDP, we optimize it with policy gradient:
\vspace{-5pt}
\begin{equation}
    \mathcal{L}_{h} = - \sum_{t=1}^{T} \hat{r}_{t} \log p^c_t,
\end{equation}
where $p_t^c$ is the probability of selecting concept $c_t$ in high-level module. $\hat{r}_{t}$ the accumulated reward of selecting $c_t$ in long term, and 
\begin{equation}\label{gamma}
    \hat{r}_t = r_t + \gamma \hat{r}_{t+1},
\end{equation}
in which $r_t$ is the learning effect (defined in Eq.(\ref{eq:learn-effect})) of the student after learn the recommended questions. \textit{i.e}, $r_T = \Delta_u$. $\gamma$ denotes the discount factor.

As the low-level module is also defined as MDP, we also optimize it with the policy gradient:
\vspace{-8pt}
\begin{equation}
    \mathcal{L}_{l} = - \sum_{t=1}^{T} \hat{r}_{t} \log p^q_t,
\end{equation}
where $p_t^q$ is the probability of selecting question $q_t$ from the question candidate set in the low-level module. 


As the learning history plays an important role in both modules, following previous work\cite{chen2023set}, we feed the representation of learning history to predict the student's proficiency in answering questions. Specifically, we feed $\textbf{h}_t$ to a binary classifier to predict the correctness of the student's response and optimize the representation according to the predicted result on ground truth:
    \begin{align}
        \hat{y}_{t} &= \delta (f_{p}(\textbf{h}_t)), \\
        \mathcal{L}_p &= - \sum_{t=1}^{T} \left( y_{t} \ln \hat{y}_{t} + (1 - y_{t}) \ln (1 - \hat{y}_{t}) \right) ,   
    \end{align}
where $\delta$ denotes the Sigmoid function and $f_p$ represents the multi-layer perceptron (MLP)\cite{taud2018multilayer}. $y_{t}$ denotes the ground-truth which indicates whether the student has correctly answered the question.

Therefore, HierLLM is optimized by 
\begin{equation}\label{alpha}
    \mathcal{L} = \mathcal{L}_h + \mathcal{L}_l + \alpha \mathcal{L}_p,
\end{equation}
here $\mathcal{L}_p$ is computed in low-level module. $\alpha$ is the hyperparameter.

\vspace{-6pt}
\section{Experiment}

In this section, we first introduce the experimental setup, followed by a detailed discussion of experiment results.
\vspace{-6pt}
\subsection{Datasets and Simulators}
\subsubsection{Datasets.}
We use two public datasets ASSIST09\footnote{https://sites.google.com/site/assistmentsdata/home/2009-2010-assistment-data}\cite{feng2009addressing} and Junyi\footnote{https://www.kaggle.com/datasets/Junyiacademy/
learning-activity-public-dataset-by-Junyi-academy} \cite{chang2015modeling} to evaluate the performance of HierLLM. The dataset provides the correspondence between concepts and questions, and students' question-answering records. Each record contains a student's ID, question ID, the correctness of the student's answer to the question, session ID, and timestamp. Records with the same session ID indicate that they originate from the same session. Within a single session, we can obtain the complete learning history of the student by organizing the records according to the timestamps. Table \ref{tab:dataset} provides statistical information about these datasets. Note that in the datasets, questions and concepts are not one-to-one correspondences; one concept can correspond to multiple questions, and one question can cover to multiple concepts.

\newsavebox{\tablebox}
\renewcommand\arraystretch{1.3}

\begin{table}[t]

    \centering
    \vspace{-5pt}
    \caption{The dataset.}
    \vspace{-8pt}

    \label{tab:dataset}

    \scriptsize

    \begin{lrbox}{\tablebox}
    \hspace{-10pt}
    \begin{tabular}{c|c|c|c|c}
     \hline
        \textbf{Dataset} & Students & Concepts & Questions & Records   \\
            \hline

        \textbf{ASSIST09} & 2968 & 121 & 15003 & 185110 \\
            \hline

        \textbf{Junyi} & 10000 & 39 & 2163 & 882198 \\
            \hline
        
	\end{tabular}
    \end{lrbox}

    \scalebox{0.95}{\usebox{\tablebox}}

\end{table}
\vspace{-3pt}
\subsubsection{Simulators.} 
Since both datasets are static and only provide a limited set of pre-collected student exercise data, which does not meet the needs of evaluation, 
we develop simulators following previous works \cite{kubotani2021rltutor, chen2023set,liu2019exploiting}.
The simulator is designed to emulate students' learning behavior (\textit{i.e.}, their performance on specific questions). Among these, \(E_a\), \(E_b\), and \(E_{\text{max}}\) in Eq.(\ref{eq:learn-effect}) are obtained from the simulator's emulation of the students' answering processes. Specifically, we designed two types of simulators:
\begin{itemize}[left=0cm]
    \item \textbf{Rule-based Simulator:} We used the KSS simulator constructed in \cite{liu2019exploiting}. In the KSS , there are a total of 10 questions, and the relationship between the questions and concepts is one-to-one. For the KSS simulator, the maximum recommendation length is 30. To model the relationship between students' learning levels and the questions, the KSS simulator employs Item Response Theory (IRT)\cite{lord1952theory, lord2012applications}, which is widely used in the field of education. IRT assumes that test questions are unidimensional and assigns a static learning state to each student. 

    \item \textbf{Knowledge Tracing(KT)-based Simulator:} The deep knowledge tracing model proposed in \cite{piech2015deep} models students' learning levels using Recurrent Neural Network(RNN)\cite{grossberg2013recurrent} to track students' evolving learning state over time and predict their feedback on given questions. This simulator is data-driven. Specifically, we trained two models, DKT \cite{shi2020learning} and IEKT \cite{long2021tracing}, using the ASSIST09 dataset and the Junyi dataset, resulting in four simulators: D-A, D-J, I-A, and I-J. For each simulator, we randomly initialized students' learning histories and learning targets. The initial learning history length is set to 20, the number of learning targets is 400, and the maximum recommendation length is 200.

\end{itemize}


\vspace{-11pt}
\subsection{Baselines} 
To evaluate the performance of our model, We compare our method with three groups of approaches. The first group consists of non-RL recommendation methods:
\begin{itemize} [left=0cm]
    \item  \textbf{Random:}
    This method randomly selects questions from the candidate set to recommend to the student.

    \item  \textbf{GRU4REC\cite{hidasi2015session}:}
    The model uses the student's learning history as input and outputs a probability distribution over the next question to be learned.

    \item  \textbf{FMLP\cite{zhou2022filter}:}
    FMLP is a sequence-based recommendation model based on a pure Multi-Layer Perceptron (MLP) architecture.

    \item  \textbf{GPT-3.5\cite{brown2020language}:}
    We used a prompt to transform the recommendation task into a natural language task and evaluated the model across five simulators without fine-tuning GPT-3.5.

\end{itemize}
The second group consists of classic RL methods:
\begin{itemize}[left=0cm]
    \item  \textbf{DQN\cite{mnih2013playing}:}
    Deep Q-Network (DQN) combines Q-learning with deep neural networks for complex decision-making.

    \item  \textbf{AC\cite{sutton1999policy}:}
    In the Actor-Critic (AC) method, the Actor recommends questions based on the student's learning state, and the Critic provides a Q-value to refine the Actor’s recommendations.

    \item  \textbf{SAC\cite{haarnoja1861soft}:}
    SAC combines policy optimization with value evaluation and adds an entropy term to promote policy exploration.

    \item  \textbf{TD3\cite{fujimoto2018addressing}:}
    TD3 enhances policy optimization stability and accuracy with clipped double Critic networks, delayed updates, and target policy smoothing.

\end{itemize}
The final group consists of RL-based question recommenders:
\begin{itemize}[left=0cm]
    \item  \textbf{CSEAL\cite{liu2019exploiting}:}
    CSEAL introduces an adaptive learning framework that can sequentially recommend appropriate questions to different learners.

    \item  \textbf{SRC\cite{chen2023set}:}
    SRC frames path recommendation as a sequence-to-sequence problem and uses a concept-aware encoder to capture correlations between learning concepts.

    \item  \textbf{GEHRL\cite{li2023graph}:}
    GEHRL introduces a hierarchical reinforcement learning framework with graph enhancement to improve target utilization in question recommendation.


\end{itemize}

\vspace{-10pt}
\subsection{Experiment Setting} 
During the training process, we fine-tune LLaMA2-7B with LoRA, the rank of LoRA is 8, the scaling parameter is 16, dropout is set to 0, and modules of LLM being fine-tuned are ["q\_proj", "v\_proj"]. In our method, we set \(d_a\) = \(d_z\) = 64, \(d_h\) = 128, \(d_m\) = 4096. The head of attention is 1 for all the cases. The $\alpha$ is 1 for all cases in Eq.(\ref{alpha}). $\gamma$ = 1 in Eq.(\ref{gamma}).
It is important to note that since KSS designed by the previous work\cite{liu2019exploiting} regards concept and question as indistinguishable, we remove the high-level module when evaluating the performance of HierLLM.
All experiments are conducted on NVIDIA A40 devices with 40GB of GPU memory. We use the Adam optimizer \cite{kingad2015methodforstochasticoptimization} with learning rates selected from $\{1\times10^{-3}, 5\times10^{-4}, 1\times10^{-4}\}$. For the parameters of the baselines, we have carefully optimized them to ensure they perform at their best.

\vspace{-7pt}
\subsection{Evaluation Metrics}
Since the goal of the question recommendation is to recommend suitable questions to students to improve their learning and accelerate their mastery of learning targets, following previous works \cite{chen2023set, liu2019exploiting, li2023graph} we apply the learning effect defined in Eq.(\ref{eq:learn-effect}) to evaluate the performance of methods. Specifically, given an arbitrary simulated student, we apply a simulator to evaluate the number of questions the student correctly answered before and after the recommendation session (\textit{i.e.}, $E_a$ and $E_b$ in Eq.(\ref{eq:learn-effect}). Then, we compute the learning effect of the student by Eq.(\ref{eq:learn-effect}). Finally, we average the learning effect of all the simulated students to evaluate the performance of recommendation methods.


\renewcommand\arraystretch{1.2}
\begin{table*}[h]
    \centering
\vspace{-6pt}
    \caption{Learning effects of different methods on 5 simulators, each result tested five times. \textbf{Bold} indicates the best performance among all methods. \underline{Underline} indicates the second-best performance. * indicates p-value < 0.05 in the significance test.}
    \vspace{-10pt}
    \label{tab:effect}
    \scriptsize
    \begin{lrbox}{\tablebox}

    \setlength{\tabcolsep}{1.2mm} {
    \begin{tabular}{c|c|
    c c c c|
    c c c c|
    c c c|c}
     \hline

        \multicolumn{2}{c|}{\multirow{2}*{\textbf{ }}} &
        \multicolumn{4}{c|}{\textbf{non-RL}} &
        \multicolumn{4}{c|}{\textbf{classic RL}} &
        \multicolumn{3}{c|}{\textbf{RL-based recommender}} &
        \textbf{ours}   \\
            \cline{3-14}
            
        \multicolumn{2}{c|}{\multirow{2}*{\textbf{ }}} & 
        Random & GRU4REC & FMLP & GPT-3.5 &
        DQN & AC & SAC & TD3 &
        CSEAL & SRC & GEHRL &
        HierLLM\\
            \hline
            
        \multirow{2}*{
        \textbf{\rotatebox{90}{KSS}}
        } &
        $t=10$ &
        0.0668$\pm$0.02 & 0.0033$\pm$0.00 & 0.0023$\pm$0.00 & 0.0150$\pm$0.01 &
        0.0793$\pm$0.04 & 0.0263$\pm$0.01 & 0.0482$\pm$0.06 & 0.0353$\pm$0.04 &
        \underline{0.1558$\pm$0.01} & 0.1157$\pm$0.02 & 0.1528$\pm$0.02 & \textbf{0.2262$\pm$0.00}* \\
            \cline{2-14}
            
        & $t=30$ &
        0.1901$\pm$0.01 & 0.0006$\pm$0.00 & 0.0024$\pm$0.00 & 0.0468$\pm$0.02 &
        0.1469$\pm$0.08 & 0.0388$\pm$0.01 & 0.0788$\pm$0.10 & 0.0848$\pm$0.10 &
        \underline{0.4835$\pm$0.02} & 0.2416$\pm$0.03 & 0.3173$\pm$0.02 & \textbf{0.4937$\pm$0.00} \\
            \hline \hline

        \multirow{3}*{
        \textbf{\rotatebox{90}{I-J}}
        } &
        $t=10$ &
        0.0070$\pm$0.03 & -0.0121$\pm$0.02 & -0.0525$\pm$0.00 & -0.0155$\pm$0.18 &
        \underline{0.1593$\pm$0.14} & -0.0085$\pm$0.02 & -0.0048$\pm$0.15 & -0.0162$\pm$0.04 &
        -0.0106$\pm$0.02 & 0.0317$\pm$0.04 & 0.0611$\pm$0.09 & \textbf{0.2905$\pm$0.00} \\
            \cline{2-14}
            
        & $t=30$ &
        -0.0009$\pm$0.03 & 0.0042$\pm$0.02 & -0.0733$\pm$0.00 & -0.0621$\pm$0.22 &
        \underline{0.2487$\pm$0.25} & -0.0032$\pm$0.02 & 0.0112$\pm$0.21 & -0.0413$\pm$0.08 &
        0.0245$\pm$0.02 & 0.0672$\pm$0.06 & 0.1357$\pm$0.20 & \textbf{0.4438$\pm$0.00} \\
            \cline{2-14}

        & $t=200$ &
        0.0307$\pm$0.02 & 0.1306$\pm$0.03 & -0.0871$\pm$0.00 & -0.0301$\pm$0.08 &
        \underline{0.4795$\pm$0.24} & 0.1548$\pm$0.15 & 0.1087$\pm$0.12 & -0.0115$\pm$0.07 &
        0.2687$\pm$0.02 & 0.2089$\pm$0.06 & 0.3130$\pm$0.30 & \textbf{0.6470$\pm$0.00} \\
            \cline{1-14}

        \multirow{3}*{
        \textbf{\rotatebox{90}{D-J}}        
        } &
        $t=10$ &
        0.0008$\pm$0.00 & 0.0010$\pm$0.00 & -0.0024$\pm$0.00 & -0.0108$\pm$0.02 &
        \underline{0.0077$\pm$0.01} & -0.0088$\pm$0.01 & 0.0069$\pm$0.02 & -0.0006$\pm$0.01 &
        -0.0005$\pm$0.00 & 0.0009$\pm$0.00 & 0.0000$\pm$0.00 & \textbf{0.01823$\pm$0.00}* \\
            \cline{2-14}
            
        & $t=30$ &
        0.0014$\pm$0.00 & 0.0000$\pm$0.00 & -0.0030$\pm$0.00 & -0.0128$\pm$0.01 &
        \underline{0.0109$\pm$0.01} & -0.0079$\pm$0.01 & 0.0063$\pm$0.02 & -0.0003$\pm$0.01 &
        -0.0004$\pm$0.00 & 0.0006$\pm$0.00 & 0.0018$\pm$0.00 & \textbf{0.01938$\pm$0.00} \\
            \cline{2-14}

        & $t=200$ &
        0.0015$\pm$0.00 & 0.0040$\pm$0.00 & -0.0030$\pm$0.00 & -0.0019$\pm$0.00 &
        \underline{0.0096$\pm$0.01} & -0.0097$\pm$0.01 & 0.0049$\pm$0.02 & 0.0040$\pm$0.00 &
        0.0014$\pm$0.00 & 0.0028$\pm$0.00 & 0.0027$\pm$0.01 & \textbf{0.01758$\pm$0.00}* \\
            \cline{1-14}

        
        \multirow{3}*{
        \textbf{\rotatebox{90}{I-A}}         
        } &
        $t=10$ &
        0.0570$\pm$0.01 & 0.0508$\pm$0.02 & -0.0078$\pm$0.00 & -0.0062$\pm$0.11 &
        0.1574$\pm$0.19 & 0.1697$\pm$0.17 & \underline{0.2205$\pm$0.19} & -0.0077$\pm$0.16 &
        -0.0217$\pm$0.04 & 0.0618$\pm$0.02 & 0.0337$\pm$0.01 & \textbf{0.3533$\pm$0.00} \\
            \cline{2-14}
            
        & $t=30$ &
        0.1309$\pm$0.02 & 0.1003$\pm$0.01 & -0.0504$\pm$0.00 & 0.0792$\pm$0.27 &
        0.2904$\pm$0.26 & 0.2648$\pm$0.20 & \underline{0.3358$\pm$0.27} & -0.0329$\pm$0.46 &
        -0.0606$\pm$0.02 & 0.1411$\pm$0.04 & 0.1269$\pm$0.02 & \textbf{0.5812$\pm$0.00} \\
            \cline{2-14}

        & $t=200$ &
        0.2734$\pm$0.02 & 0.0953$\pm$0.04 & -0.1036$\pm$0.00 & -0.0651$\pm$0.11 &
        0.5176$\pm$0.24 & 0.4127$\pm$0.17 & \underline{0.5683$\pm$0.15} & 0.1108$\pm$0.68 &
        0.1419$\pm$0.10 & 0.3041$\pm$0.01 & 0.2984$\pm$0.05 & \textbf{0.7848$\pm$0.00}* \\
            \cline{1-14}

        
        \multirow{3}*{
        \textbf{\rotatebox{90}{D-A}}        
        } &
        $t=10$ &
        0.0001$\pm$0.00 & -0.0052$\pm$0.01 & -0.0544$\pm$0.00 & -0.0541$\pm$0.05 &
        \underline{0.0150$\pm$0.01} & 0.0055$\pm$0.01 & 0.0073$\pm$0.01 & 0.0066$\pm$0.02 &
        -0.0360$\pm$0.01 & 0.0004$\pm$0.00 & -0.0034$\pm$0.01 & \textbf{0.04983$\pm$0.00}* \\
            \cline{2-14}
            
        & $t=30$ &
        0.0012$\pm$0.00 & -0.0053$\pm$0.00 & -0.0730$\pm$0.00 & -0.0811$\pm$0.05 &
        \underline{0.0175$\pm$0.02} & 0.0066$\pm$0.01 & 0.0073$\pm$0.01 & 0.0076$\pm$0.02 &
        -0.0436$\pm$0.00 & 0.0015$\pm$0.00 & 0.0058$\pm$0.01 & \textbf{0.0509$\pm$0.00}* \\
            \cline{2-14}

        & $t=200$ &
        0.0000$\pm$0.00 & -0.0020$\pm$0.00 & -0.0743$\pm$0.00 & -0.0649$\pm$0.02 &
        0.0082$\pm$0.01 & 0.0046$\pm$0.01 & \underline{0.0140$\pm$0.01} & 0.0124$\pm$0.02 &
        -0.0096$\pm$0.00 & 0.0021$\pm$0.00 & 0.0021$\pm$0.00 & \textbf{0.0510$\pm$0.00}* \\
            \hline
        
	\end{tabular}  }
    \end{lrbox}
    \scalebox{1.0}{\usebox{\tablebox}}
\vspace{-5pt}
\end{table*}

\vspace{-8pt}
\subsection{Experiment Result} \label{sec: main_result}
As KSS is designed with 30 maximum recommendation steps, we test the learning effect of simulated students within 10, 30 steps. For the KT-based simulator: D-J, D-A, I-J, I-A, we test the learning effect of simulated students within 10, 30, 200 steps. Table \ref{tab:effect} presents the performance of different methods across five simulators, each result is the average of five runs. From these results, we can observe that: 


\begin{itemize}  [left=0cm]
    \item In all cases, our HierLLM outperforms all baselines, demonstrating the effectiveness of our approach.

    \item In the majority of cases, as the number of recommendation steps increases, the model’s performance improves, meaning that students' learning effect increases. This phenomenon is reasonable as an increase in recommendation steps implies a greater number of questions exposed to the students, providing more learning opportunities to improve the learning effect. This illustrates the advantage of long-term recommendation.

    \item RL-based methods perform better than non-RL methods. That is because RL methods adjust the recommendation policy based on student feedback to adapt to the learning states of different students,
    exhibiting stronger adaptability inquestion recommendation.
    This indicates that modeling personalized problem recommendation as a Markov Decision Process (MDP) is beneficial.

    \item The performance based on GPT-3.5 is suboptimal due to the fact that in the recommendation process, the data such as questions, concepts are represented with corresponding ID, which lose most semantic information. GPT-3.5, which is trained with large number of corpus and lacks expertise in the domain of personalized question recommendation, is deficient in processing such data. 
    Therefore, targeted fine-tuning of LLMs is necessary to improve their performance in specific domains.
\end{itemize}

\renewcommand\arraystretch{1.2}\label{cold}
\begin{table}[t]
    \centering
\vspace{-6pt}
    \caption{Learning effects $\Delta_u$ under cold start problem.}
    \vspace{-10pt}
    \label{tab:effect}
    \scriptsize
    \begin{lrbox}{\tablebox}
    \setlength{\tabcolsep}{1.2mm} {
    \begin{tabular}{c|c|
    c c c c || c}
     \hline
            
        \multicolumn{2}{c|}{\multirow{2}*{\textbf{ }}} & 
        DQN & CSEAL & SRC & GEHRL & HierLLM\\
            \hline
            
        \multirow{2}*{
        \textbf{\rotatebox{90}{KSS}}
        } &
        $t=10$ &
        0.0718$\pm$0.03 & 0.1601$\pm$0.00 & 0.1242$\pm$0.01 & \underline{0.1602$\pm$0.01} & \textbf{0.2256$\pm$0.02}\\
            \cline{2-7}
            
        & $t=30$ &
        0.1273$\pm$0.01 & 
        \underline{0.4295$\pm$0.00 }& 0.1937$\pm$0.02 & 0.3290$\pm$0.01 &\textbf{0.4846$\pm$0.01} \\
            \hline \hline

        \multirow{3}*{
        \textbf{\rotatebox{90}{I-J}}
        } &
        $t=10$ &
        \underline{0.1733 $\pm$0.05} & -0.0063$\pm$0.00 & 
        0.0626$\pm$0.02 & 
        0.0315$\pm$0.01
        &\textbf{0.2098$\pm$0.00}\\
            \cline{2-7}
            
        & $t=30$ &
         0.1620$\pm$0.08 & 0.0401$\pm$0.00 & 0.1256$\pm$0.04 & 
         \underline{0.1976$\pm$0.08}
         &\textbf{0.3779$\pm$0.00}\\
            \cline{2-7}

        & $t=200$ &
        0.2521 $\pm$ 0.23 & 0.1393$\pm$0.00 & 0.2346$\pm$0.01 & \underline{0.2710$\pm$0.15}
         &\textbf{0.5838$\pm$0.00}\\
            \cline{1-7}

        \multirow{3}*{
        \textbf{\rotatebox{90}{D-J}}        
        } &
        $t=10$ &
        0.0028$\pm$0.00 & -0.0015$\pm$0.00 & 0.0002$\pm$0.00 & \underline{0.0072$\pm$0.00}
        &\textbf{0.0090$\pm$0.00}\\
            \cline{2-7}
            
        & $t=30$ &
        0.0017$\pm$0.00 & -0.0001$\pm$0.00 & 
        0.0018$\pm$0.00 &
        \underline{0.0078$\pm$0.01}
        &\textbf{0.0096$\pm$0.00}\\
            \cline{2-7}

        & $t=200$ &
        \underline{0.0085$\pm$0.00} & 
        -0.0092$\pm$0.00 & 
        0.0002$\pm$0.00 & 
        0.0032$\pm$0.01
        &\textbf{0.0103$\pm$0.00}\\
            \cline{1-7}

        \multirow{3}*{
        \textbf{\rotatebox{90}{I-A}}         
        } &
        $t=10$ &
        \underline{0.1375$\pm$0.06} & -0.0166$\pm$0.00 & 0.0450$\pm$0.00 & 0.0473$\pm$0.01
        &\textbf{0.3078$\pm$0.00}\\
            \cline{2-7}
            
        & $t=30$ &
        \underline{0.3151$\pm$0.03} & -0.0846$\pm$0.00 & 0.1433$\pm$0.01 & 0.0936$\pm$0.02
        &\textbf{0.4956$\pm$0.00}\\
            \cline{2-7}

        & $t=200$ &
        \underline{0.4770$\pm$0.21} & 0.2225$\pm$0.00 & 0.2732$\pm$0.02 & 0.2667$\pm$0.01 
        &\textbf{0.7817$\pm$0.00}\\
            \cline{1-7}

        \multirow{3}*{
        \textbf{\rotatebox{90}{D-A}}        
        } &
        $t=10$ &
        \underline{0.0175$\pm$0.02} & -0.0310$\pm$0.00 & -0.0003$\pm$0.00 & 0.0026$\pm$0.00
        &\textbf{0.0274$\pm$0.00}\\
            \cline{2-7}
            
        & $t=30$ &
        \underline{0.0181$\pm$0.02} & -0.0514$\pm$0.00 & 0.0004$\pm$0.00 & 0.0168$\pm$0.01 
        &\textbf{0.0281$\pm$0.00}\\
            \cline{2-7}

        & $t=200$ &
        \underline{0.0256$\pm$0.01} & -0.0211$\pm$0.00 & -0.0001$\pm$0.00 & 0.0002$\pm$0.00 
        &\textbf{0.0281$\pm$0.00}\\
            \hline
        
	\end{tabular}  }
    \end{lrbox}
    \scalebox{1.0}{\usebox{\tablebox}}
\vspace{-10pt}
\end{table}

\vspace{-10pt}
\subsection{Cold Start Study} \label{sec:cold-start}
A common issue in personalized question recommendation is the cold start, which occurs when the length of learning history is zero, making it challenging for the question recommendation model to provide satisfying recommendations. Additionally, the quality of initial recommendations can significantly impact student's subsequent learning behavior and learning effect. As we discussed previously, we aim to leverage the reasoning and contextual understanding capabilities of LLMs to address the issue. Therefore, to investigate the performance of HierLLM in tackling the cold start issue, we conduct the experiment. In the experiment, we remove the learning history of each simulated student at the intial step of recommendation. That is, the length of learnig history of simulated student is set to 0.

Table \ref{cold} presents the experiment results. We can observe that: 
\begin{itemize}[left=0cm]
    \item HierLLM consistently outperforms other models at all time steps (\(t=10\), \(t=30\), \(t=200\)). This result indicates that HierLLM can provide high-quality recommendations even in the absence of initial learning history. We attribute the performance of HierLLM in Table \ref{cold} to its underlying LLM, whose powerful reasoning capability and strong knowledge background supports the effective recommendations during the data-scarce cold start phase. 

    \item As the number of recommendation steps increases, the performance of HierLLM continues to improve, demonstrating its capability for incremental optimization throughout the recommendation process. Whether at shorter time steps (e.g., \(t=10\)) or longer time steps (e.g., \(t=200\)), HierLLM maintains a consistently high level of performance. 
\end{itemize}

\begin{figure}[t]
\centering
\
\includegraphics[width=1\columnwidth]{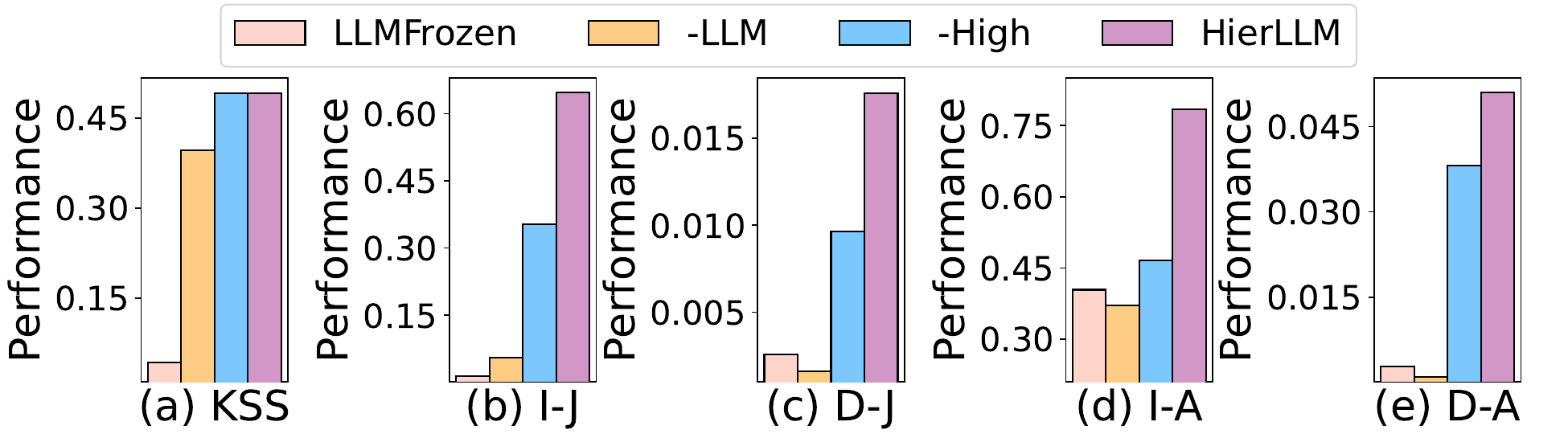} 
\vspace{-15pt}
\caption{Ablation Study}
\vspace{-19pt}

\label{fig: Ablation Study}

\end{figure}
\vspace{-10pt}
\subsection{Ablation Study} 
To further investigate the impact of modules and mechanisms on the performance, we conduct the ablation study. In the study, we have three variants:
\begin{itemize} [left = 0cm]
    \item \textbf{-High} removes the high-level module, which indicates that the hierarchical structure used for deciding the candidate set is removed. Therefore, all questions in the questions set are treated as elements of the question candidates in the low-level module.

    \item \textbf{-LLM} replace the LLM in both high-level with regular MLP. which indicates that the model no longer uses LLM for decision-making.

    \item \textbf{LLMFrozen} trains HierLLM with without fine-tuning the parameters of LLM, which indicates the parameters of LLM are always frozen. 
    
\end{itemize}




Figure \ref{fig: Ablation Study} presents the results of ablation study. We can observe from Figure \ref{fig: Ablation Study} that:
\begin{itemize}  [left=0cm]
    \item HierLLM and -High has the same performance in KSS due to the indistinguishability between concept and question.
    \item Regardless of which component is removed or replaced, the performance of HierLLM declines, indicating that all the components contribute to the performance. 
    \item HierLLM consistently surpasses that -LLM, demonstrating that LLM can effectively capture and model relationships between sequences and leverage its powerful reasoning capabilities to provide more precise and effective recommendations. Moreover, in the D-J, I-A, and D-A, LLMFrozen outperforms -LLM, which also demonstrates the strong foundational performance of LLM.
    \item 
    Without fine-tuning LLM (LLMFrozen),  the performance declines significantly, indicating that fine-tuning is essential for personalized question recommendation. 
\end{itemize}






\begin{figure}[t]
\centering
\
\includegraphics[width=1\columnwidth]{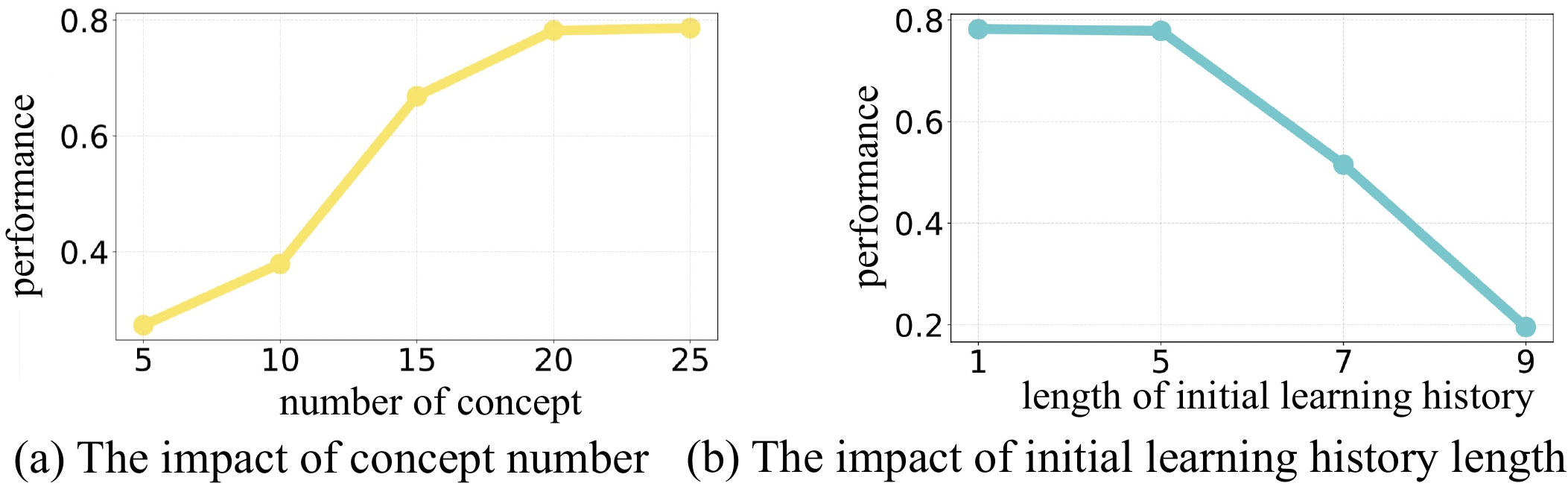} 
\vspace{-18pt}
\caption{Further analysis.}
\vspace{-18pt}

\label{fig: Hyperparameter analysis}

\end{figure}




\subsection{Further analysis} 
We investigate the impact of two factors on the performance of HierLLM:  the impact of selecting different number of concepts in high-level module and the length of initial learning history.
\vspace{-5pt}
\subsubsection{The number of concepts.}
In the high-level module, we select one concept to decrease the space for selecting recommended question. However, selecting a single concept may result in filtering out the questions that are most beneficial for students' learning at the high-level module.
Therefore, to investigate how many concepts selected in the high-level module achieve the best performance for HierLLM, we conducted further experiments and illustrate the result in Figure \ref{fig: Hyperparameter analysis}(a). We can observe that when the high-level module limits the number of selected concepts to between 1 and 5, the performance of HierLLM remains consistently high. However, when the number of concept selected by the high-level module exceeds 5, the performance begins to decline. We believe this is because, when the number of selected concepts is between 1 and 5, the number of questions in the question candidate set of the low-level module is kept within a reasonably manageable range. This results in a less complex selection space for the low-level module, making it easier to choose questions that are most beneficial for students. When the number of concept selected by the high-level module increases, the number of questions in question candidate set becomes so large that it exceeds the processing capability of HierLLM. This increases the difficulty of selecting helpful questions for students, leading to a decline in the performance.
\vspace{-4pt}
\subsubsection{The length of initial learning history.}
As we discussed in section \ref{sec:cold-start}, if the length of initial learning history is zero, the performance of models will decline. To further investigate the performance of HierLLM under different length of initial learning history, we conduct the experiment. The result is presented in Figure \ref{fig: Hyperparameter analysis}(b). We observe that the performance of HierLLM increases as the length of the initial learning history increases. Specifically, when the length increases from 5 to 10, the performance increases relatively modest. This may be due to the initial length of learning history being insufficient to fully capture the learning states of students. However, from 10 to 15, the performance increases significantly, suggesting that longer records provide more comprehensive contextual information, thereby enhancing HierLLM ability to capture the learning states of students. From 15 to 20, the rate of increase begins to slow down, likely because the model has accumulated a substantial amount of useful information, making additional records less beneficial. Beyond 20, the performance stop increasing, indicating that further records no longer significantly contribute to performance enhancement.

\begin{figure}[t]
\centering
\includegraphics[width=1\columnwidth]{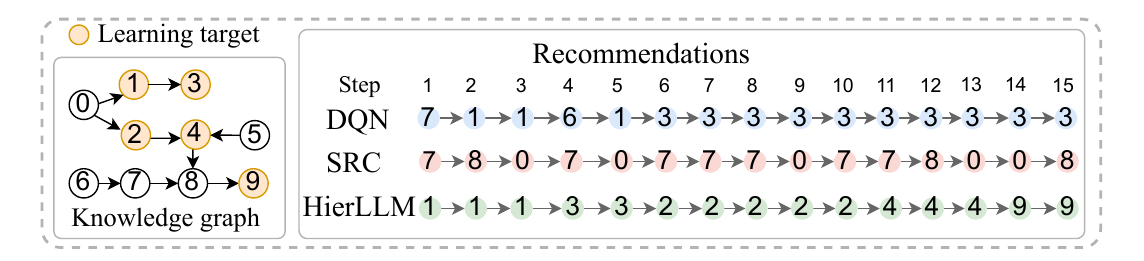} 
\vspace{-18pt}
\caption{Case Study}
\vspace{-18pt}

\label{fig: case study}

\end{figure}

\vspace{-8pt}
\subsection{Case Study}
To further investigate the interpretability of HierLLM, we apply HierLLM to recommend questions for a simulated student in KSS, as KSS has a knowledge graph to indicate the pre-requisite relation among concepts (questions), facilitating the interpretability analysis. We also selected a well-performing method from both classic RL methods (DQN) and RL-based methods (SRC), and compared them with HierLLM.
However, it is important to note that the knowledge graph is not utilized in the training of any of these methods, in order to evaluate the interpretability.
Figure \ref{fig: case study} presents the visualization, in which the learning targets are $\{1,2,3,4,9\}$ and is annotated with yellow. The recommendations are composed of 15 steps.  

From Figure \ref{fig: case study}, we can observe that the recommendations made by HierLLM fully cover the learning targets of the student and perfectly adhere to the pre-requisite relation of the knowledge graph even though the knowledge graph is not fed to HierLLM. For instance, the knowledge graph in Figure \ref{fig: case study} presents 1 is the pre-requisite concept of 3, and we can observe that 1 is recommended before 3.
In contrast, the recommendation made by DQN only covers some of 1/3 learning targets, and the recommendation made by SRC fails to cover any learning targets. However, we can observe that some of the recommendations made by both methods adhere to the pre-requisite relation of the knowledge graph even though the knowledge graph is unavailable.
Overall, the recommendations generated by HierLLM are more interpretable than those from the other two methods. We believe this is due to the extensive background knowledge embedded in the large language model (LLM) and its powerful reasoning abilities, which assist HierLLM in generating more reasonable recommendations.

\vspace{-6pt}

\vspace{-4pt}
\section{Conclusion}
In this paper, we propose a method called Hierarchical large language model for question recommendation (HierLLM) to address the temporal issue of cold start and the spatial issue of the large question set. Specifically, HierLLM is a hierarchical structure built based on LLM, leveraging the strong reasoning abilities of LLM to address the temporal issue, and leveraging the hierarchical design to address the spatial issue of the difficulty in selecting a question from the large question set. Experiment on five settings demonstrated the state-of-art performance of HierLLM, significantly enhancing the performance of long-term recommendations in the domain of personalized question recommendation. Currently, our experiments are conducted using Llama2-7B, and future work will explore the application of other LLMs for decision-making.


\bibliographystyle{ACM-Reference-Format}
\bibliography{HierLLM}
\end{document}